# Authentication via wireless networks

Darko Fudurić[1], Marko Horvat[2], Mario Žagar[3]

[1,3]Faculty of Electrical Engineering and Computing, [2]Croatian Railways Ltd.
[1,3]Unska 3, HR-10000 Zagreb, Croatia, [2]Branimirova 9a, HR-10000 Zagreb, Croatia,
E-mail: [1]darko.fuduric@fer.hr , [2]marko.horvat@hznet.hr, [3]mario.zagar@fer.hr

**Personal authentication is an important process we encounter almost every day; when we are logging on a computer, entering a company where we work, or a restricted area, when we are using our plastic credit cards to pay for a service or to complete some other financial transaction, etc. In each of these processes of personal authentication some kind of magnetic or optical token is required. But by using novel technologies like mobile computing and wireless networking, it is possible to avoid carrying multitude of ID cards or remembering a number of PIN codes. Article shows how to efficiently authenticate users via Personal Area Networks (PAN) like Bluetooth or IrDA using commonplace AES (Rijndel) or MD5 encryption. This method can be implemented on many types of mobile devices like Pocket PC PDA with Windows CE (Windows Mobile 2003) real-time operating system, or any other customized OS, so we will explain all components and key features of such basic system.**

## I. INTRODUCTION

Today small and middle range ubiquitous wireless networks allow innovative exchange of information between different parties in the communication process. In this paper we will outline a set of basic principles of using wireless technologies in controlling and restricting access to a physical object or a service.

The most important practical advantage of such approach is that only existing, and readily available communication and computer system infrastructures will be used, so there is no need for any custom computer platform or hardware. It is only required to design the appropriate software layers which will handle the authentication process.

## II. SYSTEM ARCHITECTURE

In describing any system architecture it is important to start with the most important part, and in this case that is access controlled object (ACO). ACO is a physical or software object which requires authentication in order to use its services. ACO can be cinema entrance, gallery door, company port or whatever. The important thing is that this object has wireless communication possibility (WCP), e.g. Bluetooth or IrDA. Through this WCP every mobile client device which also has WCP connects to ACO and asks for permission to use it's service or services. ACO returns an encrypted message to the mobile client which can also communicate with telecom communication infrastructure (TCI) owned by local mobile telephony provider. Mobile client communicates with TCI through any of available wireless protocols like GSM, GPRS, EDGE or UMTS. TCI has GSM/TCP-IP gateway so it is connected with custom access provider (CAP) through a standard Internet connection. CAP receives the encrypted message and provides authentication. The authentication can be successful or mobile client's request can be rejected. The entire process is illustrated in Figure 1.

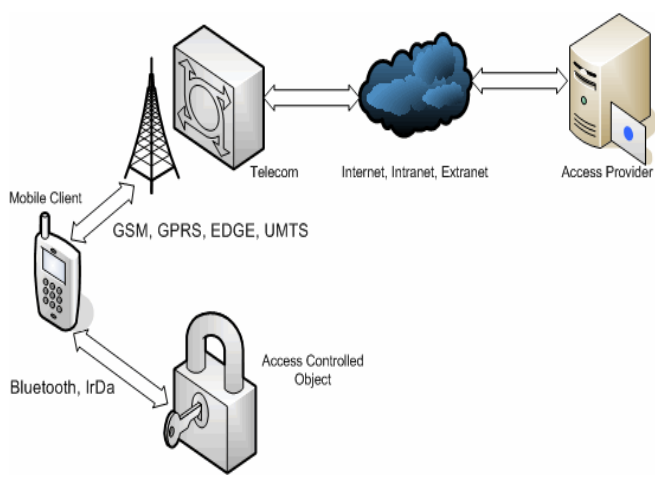

**Figure 1 – Overall system design**

## III. AUTHENTICATION OBJECT MODEL

In general, for authentication we will propose an overall object model which consists of a few processes (see Figure 2).

Authentication process starts with access request that is generated by mobile client. This request is encrypted on the service object side along with inserted unique token tag and returned to mobile client. Token tag usually describes type of service, time and other event-place information. Then, this new message passes from mobile client through telecom communication infrastructure and is received and processed by access provider, which then accepts or rejects the request according to the implemented algorithm. The algorithm itself represents behavior of a specific usage and can vary from case to case. Access provider can give only two specific responses: mobile client can get authentication, or it can not get authentication.

As it is well known, in the telecom industry there are two distinct types of client accounts: prepaid and postpaid. They differ in the way the customers pay for their wireless services, data and voice traffic. Prepaid clients pay their services in advance by purchasing a coupon or a voucher which can be used during a specific period of time, e.g. three or six months. On the other hand, postpaid clients are charged after they have consumed the telecom's service.

Taking into account these two different types of client accounts, the authentication process can fail due to four reasons which are explained in the Table 1.

Table 1 – Authentication failure use-cases

| Client Account Type | Use-case description |
|---|---|
| Prepaid | There are not enough funds on the client account |
| Prepaid | Mobile client has no access privileges to ACO |
| Postpaid | The service cannot be charged – the account is closed or error in the telecommunication's infrastructure. |
| Postpaid | Mobile client has no access privileges to ACO |

If the authentication is successful, client has access to the service and message from access provider informing client about the successful authentication process is directed back again to the event-place through TCI, service is paid and authentication is successful.

## IV. UML OBJECT MODEL AND PAYMENT PROCESS

More detailed UML explanation based on authentication and events are described in Figure 2.

Mobile client requests a key for service through telecom operator. Telecom forwards requested key to access provider which provides overall authentication process. access provider checks payment possibility and, if successful, stores result in the database and continues with returning a success message. Then, payment is executed and result of payment (sum or error) is also stored in the database. In the end, message of successful payment or error is directed back to mobile client. Client service access depends on this message, so client will temporarily save this message for later service access request.

## V. SERVICE ACCESS PROCESS

When mobile client has permission for service access, it can use this service. Figure 3 shows service access process starting with service access request. This is message from mobile client to access controlled object that provides service. Service access request is forwarded to service provider which makes authentication process again (in the dependency of access key sent with request) and, if successful, executes authorization process which will determined if client has access permission to requested service or not. If it has, service provider permits access to this service and gives service. Result of authentication and authorization is then stored in the database and message with permission is sent through telecom operator to mobile client. Client uses service in the way it sends service use message with data or commands/requests forwarded to service provider and stored in database, and service provider replies with answers (data) or actions. If an error occurs, all future servicing is prohibited and the appropriate error message is also stored in the database and forwarded to mobile client. Figure 4 encapsulates these processes in a single workflow.

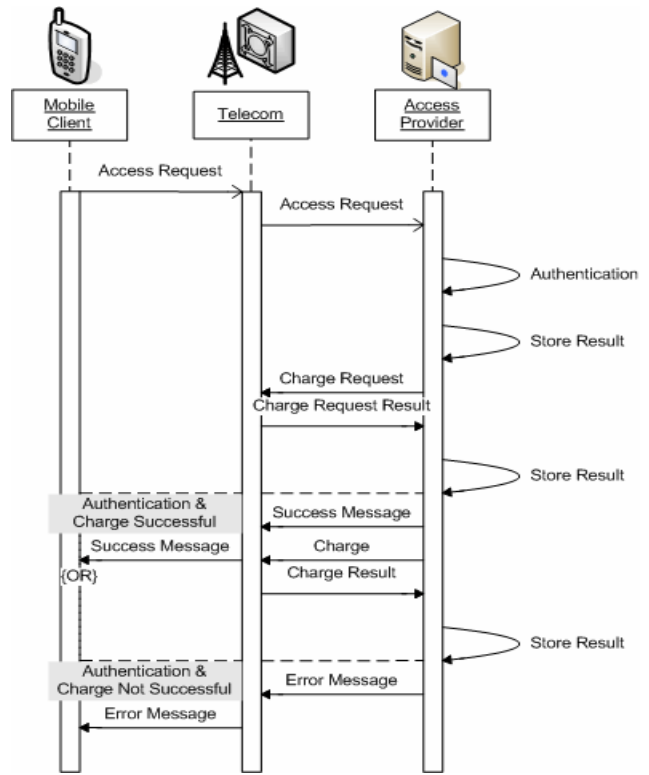

**Figure 2 - Authentication, authorization and payment processes**

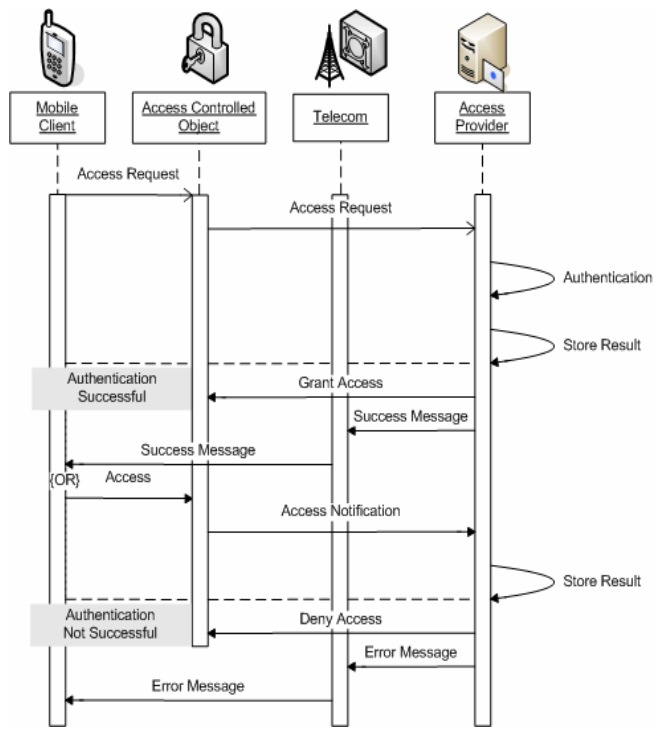

**Figure 3 - Authentication, authorization and service access processes**

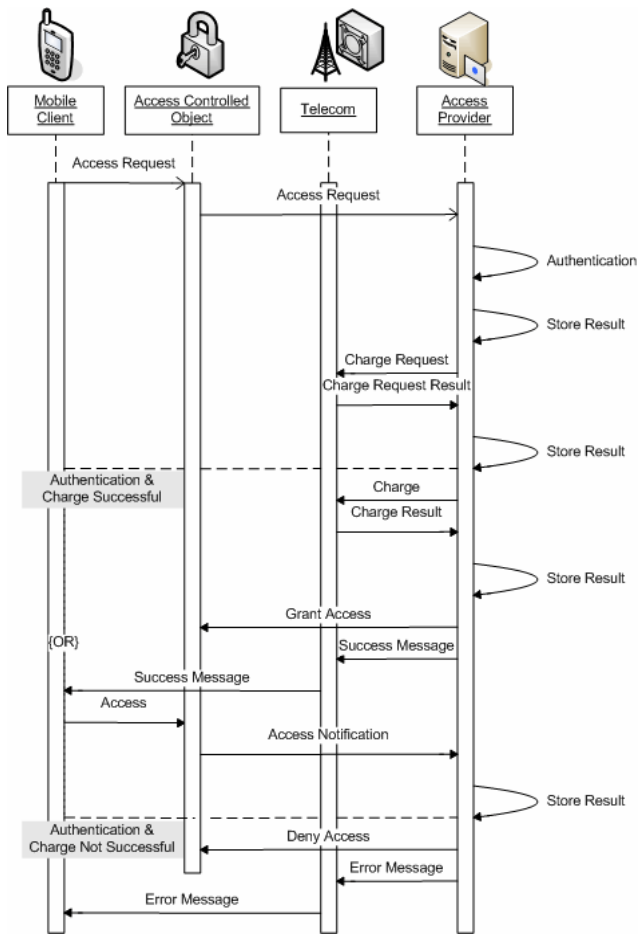

**Figure 4 - Authentication, payment and service access processing**

## VI. CRYPTOGRAPHIC ALGORITHMS

Cryptographic algorithms adequate for authorization process are MD5 and AES (Advance Encryption Standard), because of their simplicity and plenty of software implementations in Java, C/C++, JavaScript, and Microsoft .NET.

Cryptographic algorithms play a master role in authorization process because they provide privacy and access based on privileges. MD5 algorithm is usually used for storing passwords in a database. Its major advantage is fact that secret and public key are not needed, because the encrypted text can never be decrypted. Only one password is required and it is only used to encrypt the secret text. After that it is only possible to compare two encrypted messages to find out if they are equal or not. If they are equal then the secret text encrypted in them is the same. In other words, MD5 always produces the same encrypted text given the same password and secret text. The usage of MD5 encryption algorithm in authentication process has six steps as in Table 2.

AES algorithm, which is also called Rijndel, is often used for encryption and decryption of secret text messages on the client-side in multi-tier system. The algorithm has several qualities which make it desirable choice in web or mobile applications. AES is an open algorithm. It is considered safe and its software implementation is straightforward, light-weight and already ported to several program languages like JavaScript, C/C++ and others. Client and server must know the secret key for successful message exchange.

**Table 2 – Usage of MD5 encryption**

| MD5 usage algorithm: |
| --- |
| 1. User enters his password and text in the client user interface logging form.<br>2. System encrypts text with MD5 algorithm.<br>3. Encrypted text and password is then sent to database.<br>4. Encrypted text and password are compared with pre-stored encrypted text and password in database.<br>5. If they are identical, user is logged in system, otherwise encrypted text or password is wrong.<br>6. User receives the appropriate message: success or error. |

## VII. SYSTEM IMPLEMENTATION

The proposed system can be implemented with virtually any contemporary mobile technology. For example, mobile client can be any modern mobile phone that supports Java 2 Micro Edition (J2ME) or Symbian C/C++. Today all middle- or high-end cellular phones fulfill this request, even a lot of more affordable, and available, low-end models can execute custom J2ME and Symbian applications. This provides a wide base of potential client that can use this system for accessing restricted objects and services.

The same wide spectrum of choices apply to selection of access controlled object. ACO can be any industry-grade PC computer, bare-bone PC, or process computer with support for Bluetooth, GSM/GPRS/UMTS and serial networking. Depending on the specific software running on ACO, it doesn't have to have a high processing power. Direct software implementation in C or C++ and stored in the computer's read only memory (ROM) would make the most compact and perhaps the fastest solution. Of course, other types of software running on Java or Microsoft .NET are also conceivable. The most important requirement on ACO is embedded support for Personal Area Networks (PAN) and Wide Area Networks (WAN). Serial networking is required to communicate with the physical device such, e.g. doors, gates, vaults, elevators, which are access controlled and are part of ACO.

With telecom infrastructure component there are different Business-to-Business (B2B) interfaces to choose from which can be used to transfer data and information to and from the whole telecom infrastructure. This part of the system is the least defined because it greatly depends on specific implementation which varies from one to other telecom company. But since there are only a few global standards such as XML Web Services and CORBA which are always available, one can be safe to say that the problem of communicating with telecom infrastructure is always easily solvable.

Access provider is any computer system with support for B2B and high processing power that enables it to respond quickly, preferably in real-time, to all requests coming from mobile client. Access provider can be made of one or more servers, and it can be 1- or n-tier system. There are no strict limitations on access provider component as long as it provides fast, constant and reliable service.

## VIII. APPLICATION

To test the proposed concept we have decided to use only off-the-shelve technologies and rapid application design (RAD) software development tools. The goal was to create a flexible system that can be easily altered and redesigned throughout the project development process. This approach has enabled us to extensively test all involved technologies which was of the highest importance since all individual technologies are completely clear and well-defined, but the conglomeration of these technologies in a single project is a challenge.

The test system was designed around PocketPC platform and Microsoft .NET CF framework. As a mobile client we have used Asus My Pal A730W Pocket PC (Figure 5) with Windows Mobile 2003 SE operating system, Bluetooth, IrDA and Wi-Fi connectivity. Mobile client used Bluetooth to communicate with access controlled object, or Wi-Fi to gain access to the Internet and ask service provider for access key (Figure 6). ACO was an ordinary PC with Windows XP operating system and Bluetooth support. This PC emulated a door that could be opened and responded to Bluetooth messages from the mobile client. The PC had a fixed IP address. Communication with access provider, i.e. the server, was routed through LAN which emulated Internet connectivity via public GSM/GPRS networks. Therefore authentic telecom infrastructure was avoided by special software layers developed in ACO and access provider that emulated real B2B data transfer and different messages received from the telecom. Access provider was a server that listened and responded to messages from mobile client and ACO nearly in real-time. All messages were put in a queue and waited their execution according to the Figures 2 & 3. Access provider also had a fixed IP address so it could be contacted by any other three objects in the action diagram.

Our experience shows that with existing technologies this system easily can be made robust and reliable. Our intention is to move to other information platforms including those that are open-source like Linux or Linux RT operating systems and Apache web server. We will also port the mobile client's software to Symbian OS and Java 2 Micro Edition (J2ME). This will enable the system to be used on a whole range of different mobile devices not just PocketPCs. We would also like to extend GSM/GPRS communication capabilities to SMS text messages so mobile clients can use text messages to request authentication tokens and to access controlled objects. The text message would have to be of a specific predefined format and directed to the mobile telephone number of ACO or access provider. SMS messaging is a very popular form of wireless communication so by using it instead of just wireless Internet access the system could become even more appealing to the general public.

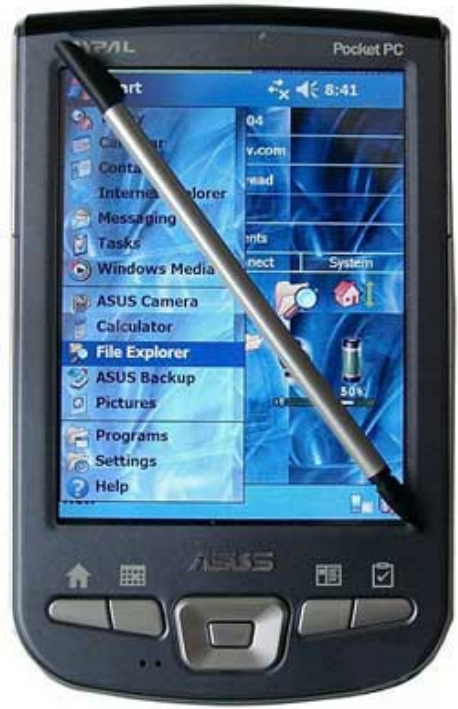

Figure 5 – Asus My Pal A730W Pocket PC as mobile client

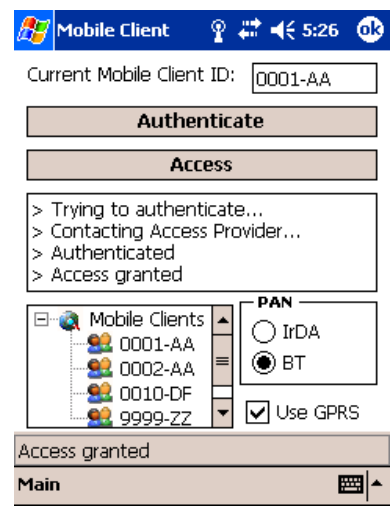

Figure 6 – Mobile client user interface

## IX. CONCLUSION

The system proposed in this article offers modern and innovative approach to the problem of controlling access to restricted object and services. By using commonplace short and middle-range wireless networking with public Internet protocols, it is possible to construct a global system that can be seamlessly introduced in any environment or to any region. For example, it is entirely possible to have only one dedicated server that controls access to hundreds of buildings and warehouses all over the world. The server can be located anywhere, without political or geographic boundaries. The system is flexible and by using virtual money it solves all problems related to payment in hard currency like monetary exchange or returning small sums of money. Also, by using off-the-shelve software platforms and existing information technology (IT) infrastructure, the whole system can be setup in very short time and with few resources. All this makes the system proposed in this article a very tempting choice as a mobile payment solution that offers numerous improvements over the traditional payment methods.

## REFERENCES

< Prored >

[1] Chakravorty, R. and Cartwright, J. and Pratt, I., "Practical Experience with TCP over GPRS", 2002.
[2] Haartsen, J., "The Bluetooth Radio System", *IEEE Personal Communications*, p. 28-36, 2000.
[3] Hansen, J. E. and Thomsen, C., "Enterprise Development with Visual Studio .NET, UML, and MSF", *Apress*, 2004.
[4] Hausmann, J.H. and Heckel, R., "Use Cases as Views: A formal approach to Requirements Engineering in the Unified Process", 2001.
[5] IRDA. IrDA Object Exchange Protocol (IrOBEX), 1997.
[6] Kalden, R. and Meirick, I. and Meyer, M., "Wireless Internet Access based on GPRS", http://www.stephan-baucke.de/publications/mme/PCM_4_2000.pdf, 2000.
[7] Praca, D. and Praden, *A-M*., "Smart cards and smart objects communication protocols: Looking to the future", 2003.
[8] Wijegunaratne, I. and Socic, M. and Chow, C., "An Architecture for Client/Server Application Software", *Australian Computer Journal*, *Vol. 26*, *No. 2*, pp. 30-41, 1994.